\documentclass[showpacs,preprintnumbers,prl,twocolumn]{revtex4}
\usepackage{amssymb}
\usepackage{amsmath}
\usepackage{graphicx}
\usepackage{bm}
\usepackage{amsfonts}

\begin{document}
\preprint{}

\title
{Qubit metrology for building a fault-tolerant quantum computer }

\author{John M. Martinis}

\email{martinis@physics.ucsb.edu}

\affiliation{University of California, Santa Barbara }
\affiliation{Google, Inc.}




\volumeyear{year}
\volumenumber{number} \issuenumber{number}
\eid{identifier}
\date{\today}

\maketitle

\textbf{Recent progress in quantum information has led to the start of several large national and industrial efforts to build a quantum computer.  Researchers are now working to overcome many scientific and technological challenges.  The program's  biggest obstacle, a potential showstopper for the entire effort, is the need for high-fidelity qubit operations in a scalable architecture.  This challenge arises from the fundamental fragility of quantum information, which can only be overcome with quantum error correction \cite{Shor95}.  In a fault-tolerant quantum computer the qubits and their logic interactions must have errors below a threshold: scaling up with more and more qubits then brings the net error probability down to appropriate levels $\sim 10^{-18}$ needed for running complex algorithms. Reducing error requires solving problems in physics, control, materials and fabrication, which differ for every implementation.  I explain here the common key driver for continued improvement - the metrology of qubit errors.}

We must focus on errors because classical and quantum computation are fundamentally different.  The classical NOT operation in CMOS electronics can have zero error, even with moderate changes of voltages or transistor thresholds.  This enables digital circuits of enormous complexity to be built as long as there are reasonable tolerances on fabrication.  In contrast, quantum information is inherently error prone because it has continuous amplitude and phase variables, and logic is implemented using analog signals.  The corresponding quantum NOT, a bit-flip operation, is produced by applying a control signal that can vary in amplitude, duration and frequency.  More fundamentally, the Heisenberg uncertainty principle states that it is impossible to directly stabilize a single qubit since any measurement of a bit-flip error will produce a random flip in phase.  The key to quantum error correction is measuring qubit parities, which detects bit flips and phase flips in pairs of qubits.  As explained in the text box, the parities are classical-like so their outcomes can be known simultaneously.

\begin{figure}[t]
\includegraphics[width=0.45\textwidth]{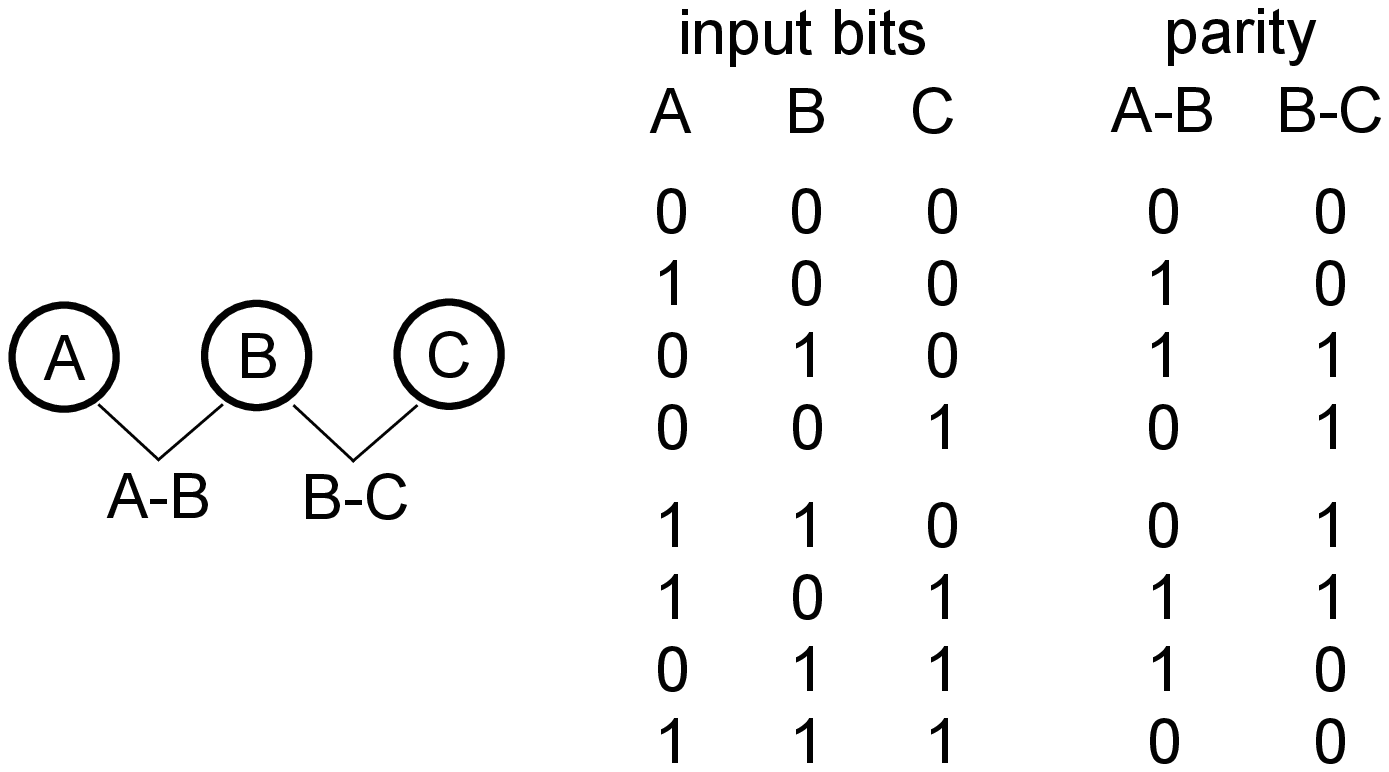}
\caption{\label{f:rep} 3-bit classical repetition code for bits A, B and C, with parity measurements between A-B and B-C.  Table shows all combination of inputs and the resulting parity measurements. For an initial state of all zeros, a unique decoding from the measurement to the actual error is obtained for only the top four entries, where there is no more than a single bit error (order $n=1$). }
\end{figure}

When parity changes, one of the two qubits had an error, but which one is not known.  To identify, encoding must use larger numbers of qubits.  This idea can be understood with a simple classical example, the 3-bit repetition code as described in Fig.\,\ref{f:rep}.  Logical states 0 (1) are encoded as 000 (111), and measurement of parities between adjacent bits A-B and B-C allows the identification (decoding) of errors as long as there is a change of no more than a single bit.  To improve the encoding to detect both order $n=1$ and $n=2$ errors, the repetition code is simply increased in size to 5 bits, with 4 parity measurements between them.  Order $n$ errors can be decoded from $2n+1$ bits and $2n$ parity measurements.

Quantum codes allow for the decoding of both bit- and phase-flip errors given a set of measurement outcomes.  As for the above example, they decode the error properly as long as the number of errors is order $n$ or less.  The probability for a decoding error can be computed numerically using a simple depolarization model that assumes a random bit- or phase-flip error of probability $\epsilon$ for each physical operation used to measure the parities.  By comparing the known input errors with those determined using a decoding algorithm, the decoding or logical error probability is found to be
\begin{align}
P_l &\simeq \Lambda^{-(n+1)} \label{eq:Pl} \\
\Lambda &= \epsilon_t/\epsilon \ ,
\end{align}
where $\epsilon_t$ is the threshold error, fit from the data.  The error suppression factor is $\Lambda$, the key metrological figure of merit that quantifies how much the decoding error drops as the order $n$ increases by one.  Note that $P_l$ scales with $\epsilon^{n+1}$, as expected for $n+1$ independent errors.  The key idea is that once the physical errors $\epsilon$ are lower than the threshold $\epsilon_t$, then $\Lambda > 1$ and making the code larger decreases the decoding error exponentially with $n$.  When $\Lambda <1$ error detection fails, and even billions of bad qubits do not help.

A key focus for fault tolerance is making qubit errors less than the threshold.  For $\Lambda$ to be as large as possible, we wish to encode with the highest threshold $\epsilon_t$.  The best practical choice is the surface code \cite{Bravyi98, Fowler12}, which can be thought of as a two-dimensional version of the repetition code that corrects for both bit and phase errors.  A $4n+1$ by $4n+1$ array of qubits performs $n$-th order error correction, where about half of the qubits are used for the parity measurements.  It is an ideal practical choice for a quantum computer because of other attributes: (i) Only nearest neighbor interactions are needed, making it manufacturable with integrated circuits. (ii) The code is upward compatible to logical gates, where measurements are simply turned off. (iii) The code is tolerant up to a significant density ($\sim 10\%$) of qubit dropouts from fabrication defects.  (iv) The high error threshold arises from the low complexity of the parity measurement; a code with higher threshold is unlikely.  (v) The simplicity of the measurement brings more complexity to the classical decoding algorithm, which fortunately is efficiently scalable.  (vi) Detected errors can be tracked in software, so physical feed-forward corrections using bit- or phase-flip gates are not needed.  (vii) The prediction Eq.\,(\ref{eq:Pl}) for $P_l$ is strictly valid only for the operative range $\Lambda \gtrsim 10$, where the threshold is $\epsilon_t \simeq 2\%$.  At break-even $\Lambda=1$, the threshold is significantly smaller $0.7\%$.

\begin{figure}[b]
\includegraphics[width=0.45\textwidth]{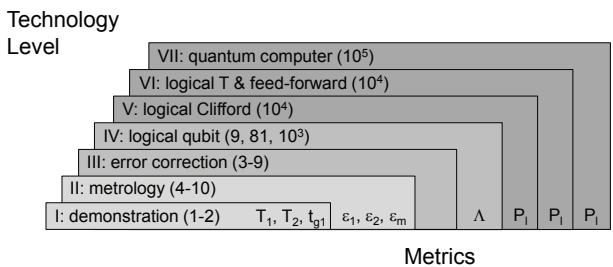}
\caption{\label{f:life} Life cycle of a qubit.  Illustration showing the increasing complexity of qubit experiments, built up upon each other, described by technology levels I through VII.  Numbers in parenthesis shows approximate qubit numbers.  Key metrics are shown at bottom.  Errors for 1 qubit, 2 qubit and measurement are described by $\epsilon_1$, $\epsilon_2$ and $\epsilon_m$, which leads to an error suppression factor $\Lambda$.  Fault-tolerant error correction is achieved when $\Lambda > 1$.  Scaling to large $n$ leads to $P_l \rightarrow 0$.  }
\end{figure}

Typical quantum algorithms use $\sim 10^{18}$ operations \cite{Fowler12}, so we target a logical error $P_l = 10^{-18}$.  Assuming an improvement $\Lambda=10$ for each order, we need  $n=17$ encoding.  The number of qubits for the surface code is $(4\cdot 17+1)^2=4761$.  For $\Lambda=100$, this number lowers by a factor of 4.  Although this seems like a large number of qubits from the perspective of present technology, we should remember that a cell phone with $10^{12}$ transistors, now routinely owned by most people in the world, was inconceivable only several decades ago.

Hardware requirements can be further understood by separating out the entire parity operation into one- and two-qubit logic and measurement components.  Assuming errors in only one of these components, break-even thresholds are respectively 4.3\%, 1.25\% and 12\%: the 2-qubit error is clearly the most important, whereas measurement error is the least important.  For the practical case when all components have non-zero errors, I propose the threshold targets
\begin{align}
\epsilon_{1}&\leq 0.1\% \\
\epsilon_{2}&\leq 0.1\% \\
\epsilon_{m}&\leq 0.5\% \ ,
\end{align}
which gives $\Lambda \geq 17$.  It is critical that all three error thresholds be met,  as the worst performing error limits the logical error $P_l$.  Measurement errors $\epsilon_{m}$ can be larger because its single component threshold 12\% is high.
Two-qubit error $\epsilon_2$ is the most challenging because its physical operation is much more complex than for single qubits.  This makes $\epsilon_2$ the primary metric around which the hardware should be optimized.  The single qubit error $\epsilon_1$, being easier to optimize, should readily be met if the two-qubit threshold is reached.  Note that although it is tempting to isolate qubits from the environment to lower one-qubit errors, in practice this often makes it harder to couple them together for two-qubit logic; I call such strategy ``neutrino-ized qubits''.

In the life cycle of a qubit technology, experiments start with a single qubit and then move to increasingly more complex multi-qubit demonstrations and metrology.  The typical progression \cite{Devoret13} is illustrated in Fig.\,\ref{f:life}, where the technology levels and their metrics are shown together.

In level I, one and two-qubit experiments measure coherence times $T_1$ and $T_2$, and show basic functionality of qubit gates.  Along with the one-qubit gate time $t_{g1}$, an initial estimate of gate error can be made.  Determining the performance of a two-qubit gate is much harder since other decoherence or control errors will typically degrade performance.  Swapping an excitation between two qubits is a simple method to determine whether coherence has changed.  Quantum process tomography is often performed on one- and two-qubit gates \cite{Benhelm08}, which is important as it proves that proper quantum logic has been achieved.  In this initial stage, it is not necessary to have low measurement errors, and data often have arbitrary units on the measurement axis.  This is fine for initial experiments that are mostly concerned with the performance of qubit gates.

In level II, more qubits are measured in a way that mimics the scale-up process.  This initiates more realistic metrology tests as to how a qubit technology will perform in a full quantum computer.  Here, the application of many gates in sequence through randomized benchmarking (RB) enables the total error to grow large enough for accurate measurement, even if each gate error is tiny \cite{Ryan09}.  Interleaved RB is useful for measuring the error probability of specific one- and two-qubit logic gates, and gives important information on error stability.  Although RB represents an average error and provides no information on error coherence between gates, it is a practical metric to characterize overall performance \cite{Barends14}.  For example, RB can be used to tune up the control signals for lower errors \cite{Kelly14b}.  Process tomography can be performed for multiple qubits, but is typically abandoned because (i) the number of necessary measurements scales rapidly with increasing numbers of qubits, (ii) information on error coherence is hard to use and (iii) it is difficult to separate out initialization and measurement errors.  Measurement error is also obtained in this level; differentiation should be made between measurement that destroys a qubit state or not, since the latter is eventually needed in level IV for logical qubits.  A big concern is crosstalk between various logic gates and measurement outcomes, and whether residual couplings between qubits create errors when no interactions are desired.  A variety of crosstalk measurements based on RB are useful metrology tools.

In level III an error detection or correction algorithm is performed \cite{Nigg14}, representing a complex systems test of all components.  Qubit errors have to be low enough to perform many complex qubit operations.  Experiments work to extend the lifetime of an encoded logical state, typically by adding errors to the various components to show improvement from the detection protocol relative to the added errors.

At level IV, the focus is measuring $\Lambda > 1$, demonstrating how a logical qubit can have less and less error by scaling up the order $n$ of error correction.  The logical qubit must be measured in first and second order, which requires parity measurements that are repetitive in time so as to include the effect of measurement errors.  Note that extending the lifetime of a qubit state in first order is not enough to determine $\Lambda$.  Measuring $\Lambda > 1$ indicates that all first order decoding errors have been properly corrected, and that further scaling up should give lower logical errors.  Because 81 qubits are needed for the surface code with $n=2$, a useful initial test is for bit-flip errors, requiring a linear array of 9 qubits.  These experiments are important since they connect the error metrics of the qubits, obtained in level II, to actual fault-tolerant performance $\Lambda$.  As there are theoretical and experimental approximations in this connection, including the depolarization assumption for theory and RB measurement for experiment, this checks the whole framework of computing fault-tolerance.  A fundamentally important test for $n \geq 2$ is whether $\Lambda$ remains constant, since correlated errors would cause $\Lambda$ to decrease. Level IV tests continue until the order $n$ is high enough to convincingly demonstrate an exponential suppression of error.  A significant challenge here is to achieve all error thresholds in one device and in a scalable design.

An experiment measuring the bit-flip suppression factor $\Lambda_X$ has been done with a linear chain of 9 superconducting qubits \cite{Kelly14a}.  The measurement $\Lambda_X = 3.2$ shows logical errors have been reduced, with a magnitude that is consistent with the bit-flip threshold of $3\%$ and measured errors.  This is the first demonstration that individual error measurements can be used to predict fault tolerance.  For bit and phase fault tolerance, we need to improve only two-qubit errors and then scale.

In level V, since the lifetime of a logical state has been extended, the goal is to perform logical operations with minuscule error.  Similar to classical logic that can be generated from the NOT and AND gates, arbitrary quantum logic can be generated from a small set of quantum gates.  Here all the Clifford gates are implemented, such as the S,  Hadamard, or controlled-NOT.  The logical error probabilities should be measured and tested for degradation during logical gates.

In level VI, the test is for the last and most difficult logic operation, the T gate, which phase shifts the logical state by $45^\circ$.  Here, state distillation must be demonstrated, and feed-forward from qubit errors conditionally controls a logical S gate \cite{Fowler12}.  Because logical errors can be readily accounted for in software for all the logical Clifford gates in level V, feed-forward is only needed for this non-Clifford logical T gate.

Level VII is for the full quantum computer.

The strategy for building a fault-tolerant quantum computer is as follows. At level I, the coherence time should be at least 1000 times greater than the gate time.  At level II, all errors need to be less than threshold, with particular attention given to hardware architecture and gate design for lowest 2 qubit error.  Design should allow scaling without increasing errors.  Scaling begins at level IV: 9 qubits give the first measurement of fault tolerance with $\Lambda_X$, 81 qubits give the proper quantum measure of  $\Lambda$, and then about $10^3$ qubits allow for exponentially reduced errors.  At level V through VII, $10^4$ qubits are needed for logical gates, and finally about $10^5$ qubits will be used to build a demonstration quantum computer.

The discussion here focuses on optimizing $\Lambda$, but having fast qubit logic is desirable to obtain a short run time.  Run times can also be shortened by using a more parallel algorithm, as has been proposed for factoring.  A 1000 times slower quantum logic can be compensated for with about 1000 times more qubits.

Scaling up the number of qubits while maintaining low error is a crucial requirement for level IV and beyond.  Scaling is significantly more difficult than for classical bits since system performance will be affected by small crosstalk between the many qubits and control lines.  This criteria makes large qubits desirable, since more room is then available for separating signals and incorporating integrated control logic and memory.  Note this differs from standard classical scaling of CMOS and Moore's law, where the main aim is to decrease transistor size.

Superconducting qubits have macroscopic wavefunctions and are therefore well suited for the challenges of scaling with control.  I expect qubit cells to be in the $30-300\,\mu$m size scale, but clearly any design with millions of qubits will have to properly tradeoff density with control area based on experimental capabilities.

In conclusion, progress in making a fault-tolerant quantum computer must be closely tied to error metrology, since improvements with scaling will only occur when errors are below threshold.  Research should particularly focus on two-qubit gates, since they are the most difficult to operate well with low errors.  As experiments are now within the fault-tolerant range, many exciting developments are possible in the next few years.

The author declares no competing financial interests.

\

\noindent\fbox{
\begin{minipage} {24em}
\textbf{Quantum parity.}  An arbitrary qubit state is written as $\Psi=\cos(\theta/2)|0\rangle+e^{i\phi}\sin(\theta/2)|1\rangle$, where the continuous variables $\theta$ and $\phi$ are the bit amplitude and phase.  A bit measurement collapses the state into $|0\rangle$ ($|1\rangle$) with probability $\cos^2(\theta/2)$ ($\sin^2(\theta/2)$), thus digitizing error.  In general, measurement allows qubit errors to be described as either bit flip $\hat{X}$ ($\,|0\rangle \leftrightarrow |1\rangle\,$) or phase flip $\hat{Z}$ ($\,|1\rangle \leftrightarrow -|1\rangle\,$).  According to the Heisenberg uncertainty principle, it is not possible to simultaneously measure the amplitude and phase of a qubit, so obtaining information on a bit flip induces information loss on phase equivalent to a random phase flip, and vice versa.  This property comes fundamentally from bit and phase flips not commuting $[\hat{X},\hat{Z}] = \hat{X}\hat{Z}- \hat{Z}\hat{X} \neq 0$; the sequence of the two operations matter.  Quantum error correction takes advantage of an interesting property of qubits $\hat{X}\hat{Z} = -\hat{Z}\hat{X}$, so that a change in sequence just produces a minus sign.  With $\hat{X_1}\hat{X_2}$ and $\hat{Z_1}\hat{Z_2}$ corresponding to 2-qubit bit and phase parities, they now commute because a minus sign is picked up from each qubit
\begin{align}
[\hat{X_1}\hat{X_2},\hat{Z_1}\hat{Z_2}] &=
\hat{X_1}\hat{X_2}\hat{Z_1}\hat{Z_2}-\hat{Z_1}\hat{Z_2}\hat{X_1}\hat{X_2} \\
&=\hat{X_1}\hat{X_2}\hat{Z_1}\hat{Z_2}-(-)^2 \hat{X_1}\hat{X_2}\hat{Z_1}\hat{Z_2} \\
&= 0 \ .
\end{align}
The two parities can now be known simultaneously, implying they are classical-like: a change in one parity can be measured without affecting the other.
\end{minipage}
}


\begin{thebibliography}{99}

\bibitem{Shor95} P. W. Shor, Phys. Rev. A \textbf{52}, R2493 (1995).

\bibitem{Bravyi98} S. B. Bravyi and A. Yu. Kitaev, arXiv quant-ph/9811052 (1998).

\bibitem{Fowler12} A. G. Fowler, M. Mariantoni, J. M. Martinis and A. N. Cleland, Phys. Rev. A \textbf{ 86}, 032324 (2012).

\bibitem{Devoret13} M. H. Devoret and R. J. Schoelkopf, Science \textbf{339}, 1169 (2013).

\bibitem{Benhelm08} J. Benhelm, G. Kirchmair, C. F. Roos and R. Blatt, Nature Physics \textbf{4}, 463 (2008).

\bibitem{Ryan09} C. A. Ryan, M. Laforest and R. Laflamme, New J. Phys. \textbf{11}, 013034 (2009).

\bibitem{Barends14} R. Barends, J. Kelly, A. Megrant, A. Veitia, D. Sank, E. Jeffrey \textit{et. al.}, Nature \textbf{508}, 500 (2014).

\bibitem{Kelly14b} J. Kelly, R. Barends, B. Campbell, Y. Chen, Z. Chen, B. Chiaro \textit{et. al.}, Phys. Rev. Lett. \textbf{112}, 240504 (2014).

\bibitem{Nigg14} D. Nigg, M. Müller, E. A. Martinez, P. Schindler, M. Hennrich, T. Monz \textit{et. al.}, Science \textbf{345}, 302 (2014).

\bibitem{Kelly14a} J. Kelly, R. Barends, A. G. Fowler, A. Megrant, E. Jeffrey, T. C. White \textit{et. al.}, Nature \textbf{519}, 66 (2015).









\end{thebibliography}
\end{document}